\documentclass[a4paper]{jpconf}
\usepackage{citesort}
\usepackage{aas_macros}
\usepackage{graphicx}
\usepackage{amssymb}

\begin{document}
\title{AGN accretion disks as spatially resolved by polarimetry}

\author{\bf Makoto Kishimoto$^{1,2}$, Robert Antonucci$^{3}$, Omer
Blaes$^{3}$, Andy Lawrence$^{2}$, Catherine Boisson$^{4}$, Marcus
Albrecht$^{5}$, Christian Leipski$^{3}$}

\address{$^1$ Max-Planck-Institut f\"ur Radioastronomie, Auf dem H\"ugel 69,
 53121 Bonn, Germany}
\address{$^2$ SUPA (Scottish Universities Physics Alliance), Institute for
 Astronomy, University of Edinburgh, Royal Observatory, Blackford
 Hill, Edinburgh, EH9 3HJ, UK}
\address{$^3$ Physics Department, University of California, Santa Barbara, CA
 93106, USA}
\address{$^4$ LUTH, FRE 2462 du CNRS, associ\'ee \`a l'Universit\'e
    Denis Diderot, Observatoire de Paris, Section de Meudon, F--92195
    Meudon Cedex, France}
\address{$^5$ Instituto de Astronom\'ia, Universidad Cat\'olica del Norte
   (UCN), Avenida Angamos 0610, Antofagasta, Chile}

\ead{mk@mpifr-bonn.mpg.de}

\begin{abstract}

A crucial difficulty in understanding the nature of the putative
accretion disk in AGNs is that some of its key intrinsic spectral
signatures cannot be observed directly.  The strong emissions from the
broad-line region (BLR) and the obscuring torus, which are generally
yet to be spatially resolved, essentially 'bury' such signatures.
Here we argue that we can actually isolate the disk emission spectrum
by using optical and near-infrared polarization of quasars and uncover
the important spectral signatures.  In these quasars, the polarization
is considered to originate from electron scattering interior to the
BLR, so that the polarized flux shows the disk spectrum with all the
emissions from the BLR and torus eliminated.  The polarized flux
observations have now revealed a Balmer edge feature in absorption and
a blue near-infrared spectral shape consistent with a specific and
robust theoretical prediction.  These results critically verify the
long-standing picture of an optically thick and locally heated disk in
AGNs.

\end{abstract}

\section{The key buried signatures from the central accretion disk}

The primary radiative output of active galactic nuclei (AGNs) is
observed at the ultraviolet/optical wavelengths. This is attributed to
be from an accretion disk around a supermassive black hole.  While
this putative accretion disk has been modeled extensively, it is well
known that there are disagreements between observations and model
predictions in a few major respects
(e.g. refs. \cite{Antonucci99,Koratkar99}).  A crucial observational
difficulty here has been that we still do not have enough spatial
resolution to isolate the accretion disk from the surrounding
regions. Important spectral features of the disk are thus often buried
under the strong emission from these regions --- in particular, from
the broad-line-emitting region (BLR) and from a slightly larger-scale
torus-like region with hot dust grains.

One such key spectral region is the near-infrared.  In the fundamental
hypothesis of the standard, most extensively studied model
\cite{Shakura73}, the disk is optically thick and heated by local
energy dissipation, and this sets the effective disk temperature $T$
as a function of radius $r$ as $T \propto r^{-3/4}$ over a broad range
of radii.  This leads to the well-known blue spectral shape at long
wavelengths, $F_{\nu} \propto \nu^{+1/3}$, in the simple case of local
blackbody emission.  In more sophisticated, bare-disk atmosphere
models (e.g. ref. \cite{Hubeny00}), the same blue limit is reached
longward of $\sim$1 $\mu$m essentially independent of parameters
suitable for quasars.  Then we should be able to robustly test disk
models here. Furthermore, the standard disk is also well known to be
gravitationally unstable in the outer radii \cite{Shlosman87}, which
may correspond to those emitting in the near-infrared
\cite{Goodman03}.  The spectrum may show a break due to a possible
disk truncation and become even bluer toward longer wavelengths. Thus
the near-infrared disk spectrum is quite important for the tests of
disk models.  However this is almost exactly where the dust thermal
emission from the torus starts to dominate the spectrum (set by dust
sublimation temperature). Thus this important near-infrared disk
spectrum usually cannot be observed.

Another key spectral signature is the Balmer edge.  Among the opacity
edge features generally predicted by disk atmosphere models, the
Balmer edge has the advantage (over the Lyman edge) of being much less
prone to foreground absorptions.  However, high-order Balmer emission
lines and Balmer continuum (and also FeII blends) from the BLR bury
the Balmer edge feature in the disk spectrum making it essentially
unobservable.

However, we argue here that we can separate the disk emission from the
surrounding emissions by using optical and near-infrared polarization
of quasars. With this measurement, effectively gaining very high
spatial resolution, we can study these key spectral signatures from
the accretion disk.

\begin{figure}[b]
\begin{center}
\includegraphics[width=12cm]{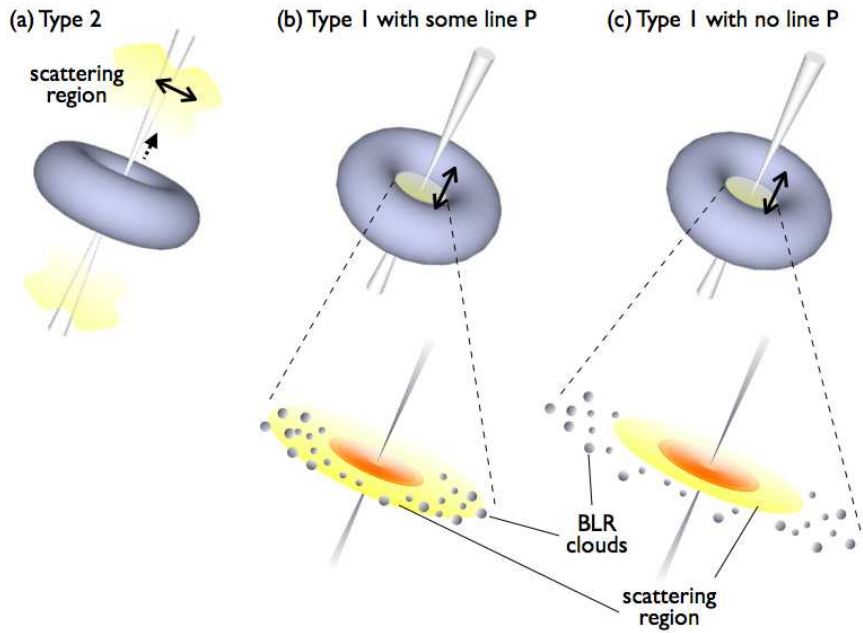}

\caption{\label{fig_type2vs1}Schematic diagrams for the geometry of
dominant scattering regions for (a) Type 2s (b) Type 1s with some line
polarization (c) Type 1s with no line polarization. In each panel, the
double arrow shows the position angle of continuum polarization.}

\end{center}
\end{figure}

\section{The optical polarization of quasars or Type 1 AGNs} 

Perhaps the most well-known optical polarization in AGNs is the one
seen in Type 2 objects, where the torus obscures our line of sight to
the accretion disk and the BLR.  The broad emission lines are seen in
the polarized flux in many of these, with polarization position angle
(PA) perpendicular to the radio jet axis \cite{Antonucci93}. The
interpretation is that the gas which resides {\it outside} the BLR,
along the jet axis above and below the torus, scatters the light from
the accretion disk {\it and} the BLR into our line of sight
(Fig.1a). Thus they both show up in the polarized flux.

Here we are interested {\it not} in these Type 2 objects, but rather
in Type 1 objects, namely Seyfert 1 galaxies and quasars.  In those
objects, our line of sight is much less inclined, giving a direct view
of the bright nuclear region interior to the torus. This gives rise to
a different nuclear polarization component to dominate. The optical
continuum is often polarized at polarization degrees $P$ of
$\lesssim$1\% level, with PAs {\it parallel} to the jet axis
(e.g. ref.\cite{Berriman90}).  In many Seyfert 1s, the broad lines are
also polarized but often at lower $P$ and at different PA than
continuum (it rotates across the line wavelengths;
e.g. ref.\cite{Smith04may}). These imply that the scattering region is
more or less similar in size to the BLR. The parallel polarization
quite possibly indicates that the scattering region is in a
flattened/equatorial optically-thin geometry having its symmetry axis
along the jet direction (Fig.1b).

At least in some quasars, however, similar continuum polarization is
seen but with no or very little line emission in the polarized
flux. In this case, scattering is considered to be caused {\it
interior} to the BLR (Fig.1c), by electrons (since the site is inside
the dust sublimation radius). Then the polarized flux would in fact be
an electron-scattered copy of the intrinsic spectrum of the central
engine, with all the emissions from the BLR and torus eliminated. This
polarized flux enables us to isolate the accretion disk spectra from
the contaminating emissions.

\begin{figure}[b]
\begin{center}
\includegraphics[width=10cm]{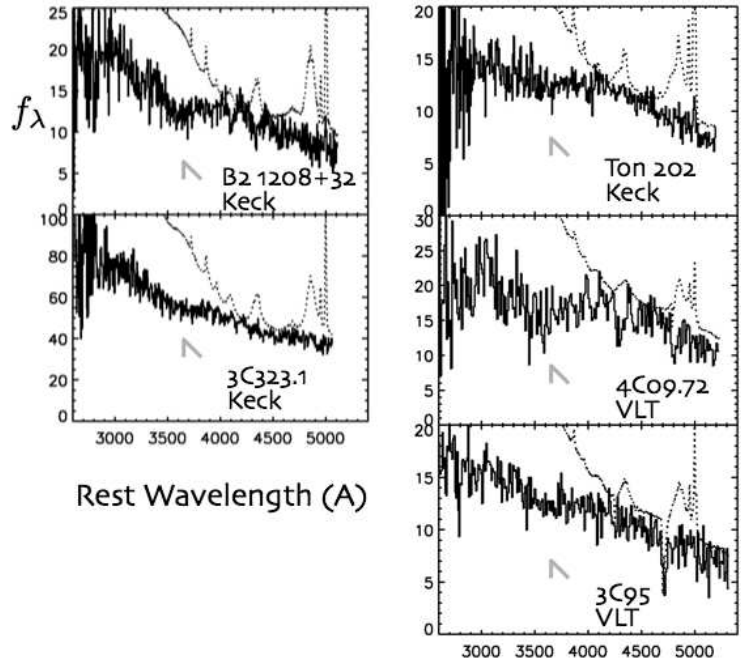}

\caption{\label{fig_balm}Optical spectropolarimetry of five quasars
from ref.\cite{Kishimoto04}. The solid line is the polarized flux in
units of 10$^{-18}$ ergs/cm$^2$/sec/\AA. The dotted line is the total
flux scaled to match the polarized flux at the red side. The
wavelength of the Balmer discontinuity, 3646\AA, is indicated as a
folded line in each panel. }

\end{center}
\end{figure}

\begin{figure}[b]
\begin{center}
\includegraphics[width=10cm]{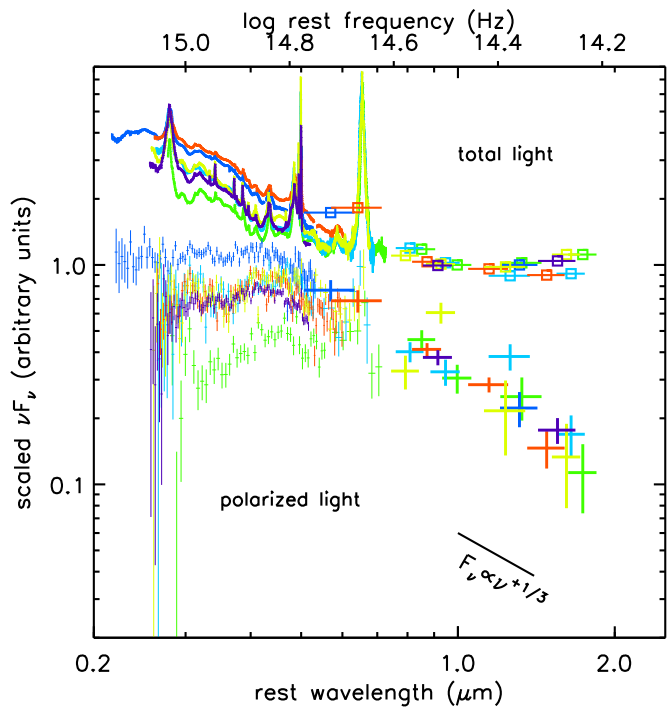}
\caption{\label{fig_nearIR}
Overlay of the polarized and total flux spectra observed in six
different quasars, from ref.\cite{Kishimoto08}. We plot scaled $\nu
F_{\nu}$ data: Q0144-3938 (redshift $z$=0.244), green; 3C95
($z=$0.616), blue; CTS A09.36 ($z$=0.310), light blue; 4C 09.72
($z$=0.433), red; PKS 2310-322 ($z$=0.337), light green; Ton 202
($z$=0.366), purple.  Total flux spectra, shown as bold traces in the
optical and as squares in the near-infrared, are normalized at 1$\mu$m
in the rest frame, by interpolation.  Polarized flux spectra, shown as
light points in the optical and as bold points in the near-infrared
(vertical error bars, 1-$\sigma$), are separately normalized, also at
1$\mu$m, by fitting a power-law to the near-infrared polarized flux spectra.
The normalized polarized flux spectra are arbitrarily shifted
downwards by a factor of three relative to the normalized total flux
spectra, for clarity.}
\end{center}
\end{figure}

\begin{figure}[t]
\includegraphics[width=10cm]{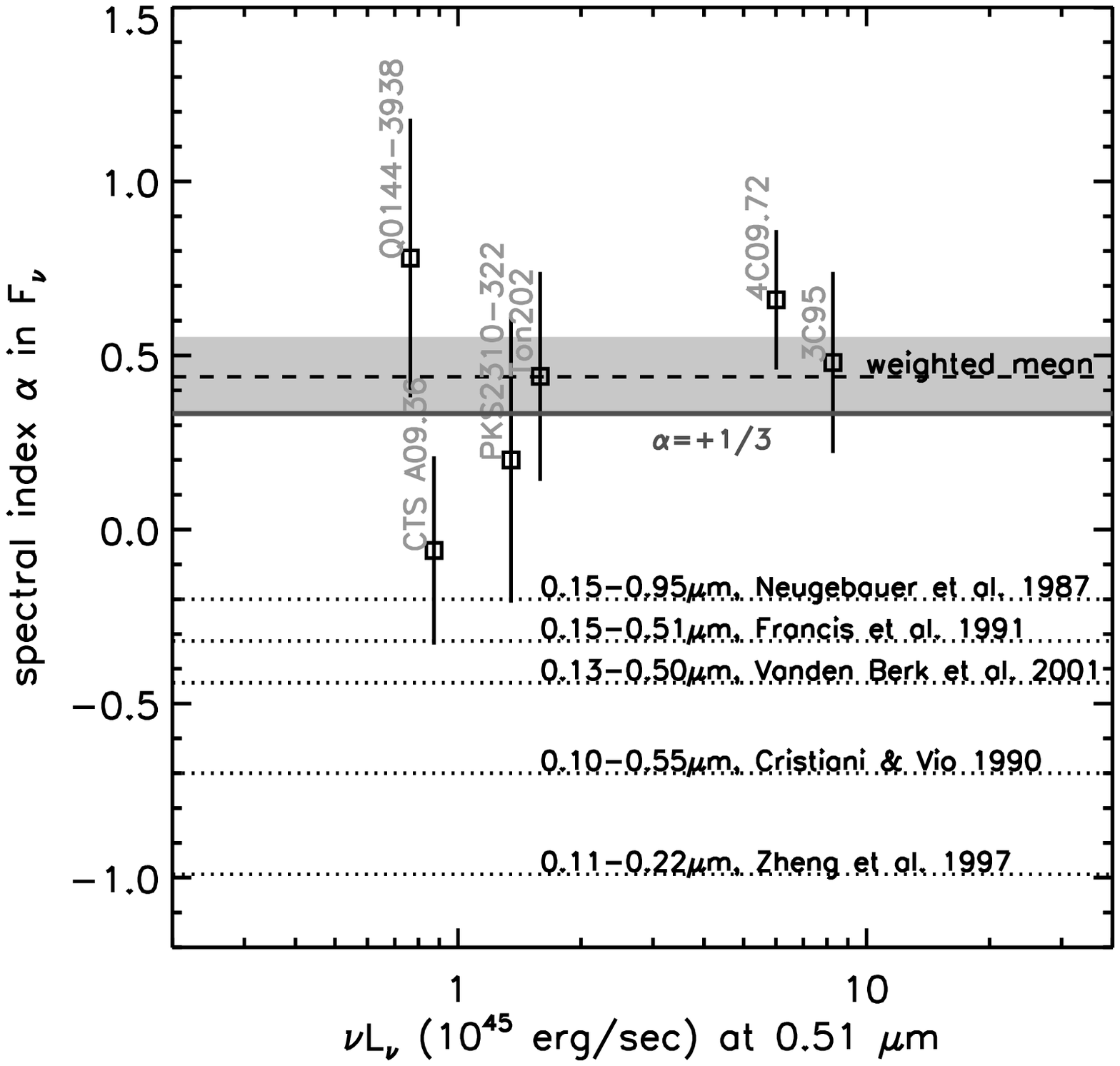}\hspace{0.5cm}%
\begin{minipage}[b]{5.5cm}\caption{\label{fig_slope}
Spectral index of polarized flux spectra, from
ref.\cite{Kishimoto08}. We plot $\alpha$ (in $F_{\nu} \propto
\nu^{\alpha}$) against $\nu L_{\nu}$ for total light at 0.51
$\mu$m. The index was measured by a power-law fit for each
near-infrared polarized flux spectrum (note the different wavelength
range covered depending on the redshift) and is shown with 1-$\sigma$
error bars. A weighted mean of the spectral index measurements is
shown dashed; the shaded area represents its deduced 1-$\sigma$
uncertainty.  The mean or median slopes of the ultraviolet/optical
total flux spectra derived in various studies
\cite{Neugebauer87,Francis91,VandenBerk01,Cristiani90,Zheng97} are
also shown.}
\end{minipage}
\end{figure}

\section{Intrinsic disk spectra as revealed by polarimetry}

Figure~\ref{fig_balm} shows the polarized flux spectra of such
quasars.  In contrast to the total flux spectra, we see essentially no
emission lines in the polarized flux spectra.  Thus the polarized flux
is very likely to show the intrinsic spectral behavior of the disk
without the BLR emission contamination. The objects were chosen to be
at redshift $\gtrsim$0.3 to make sure that the Balmer edge region is
covered with good sensitivity.  The feature is indeed seen, all in
absorption: there is a downturn at around 4000\AA\ and an upturn at
around 3600\AA\ in all the objects shown.  Thus the fundamental
implication here is that the emission is thermal and optically thick
in nature.

Then we have extended the work to longer wavelengths: since the
polarization originates interior to the dusty torus, we should also be
able to eliminate the dust emission and uncover the underlying
near-infrared disk spectrum.  Figure~\ref{fig_nearIR} shows the
results for six quasars \cite{Kishimoto05,Kishimoto08}. Some of the
quasars are those shown in Figure~\ref{fig_balm}, while the others
have newly been found to be suitable for this work (i.e. no or very
little line flux in polarized light) from our optical polarimetric
survey with the ESO3.6m telescope and follow-up spectropolarimetry
with the VLT \cite{Kishimoto08}.  In all six objects, the total flux
spectra in $\nu F_{\nu}$ show an up-turn longward of $\sim$1 $\mu$m
due to the onset of dust emission. However, the polarized flux spectra
all show systematically a rapid decrease toward long wavelengths with
a shape of approximately power-law form.  The spectral indices
$\alpha$ measured in $F_{\nu} \propto \nu^{\alpha}$ for the
near-infrared polarized flux are shown in Figure~\ref{fig_slope}, and
compared with those observed at the optical/ultraviolet wavelengths.
The uncovered near-infrared colors are clearly much bluer than those
at the shorter wavelengths. Surprisingly they are all consistent with
the shape $\nu^{+1/3}$, with an weighted mean of $\alpha =
+0.44\pm0.11$.

The systematic behavior of the polarized flux, and the fact that PAs
are observed to be essentially constant over the whole wavelengths
(from the ultraviolet to near-infrared) in each object
\cite{Kishimoto08}, strongly argue against there being any secondary
polarization component arising in the near-infrared.  Therefore, the
near-infrared polarized flux spectra are very likely to reveal the
intrinsic spectra of accretion disks.  The measured slopes, which are
as blue as the predicted spectral shape of $\nu^{+1/3}$, strongly
suggest that, at least in the outer near-infrared emitting radii, the
standard but unproven picture of the disk being optically thick and
locally heated is approximately correct. In this case, other model
problems at shorter wavelengths should be directed to the lack of our
understanding of the inner regions of the same disk.

\section{Outlook}

The optical and near-infrared polarization measurements of quasars
have turned out to be quite revealing, and have been delineating the
fundamental aspects of the accretion disk in their central engine.  A
further question is: does the revealed near-infrared spectrum show an
indication of disk truncation in this outer region? Although
statistically insignificant, the data do suggest that the slope is
slightly bluer than the shape $\nu^{+1/3}$. Theoretical modeling is
underway with the current data. Further measurements may provide
totally new insight on the outer edge of the disk and how material is
being supplied to the nucleus.

\section*{References}


\providecommand{\newblock}{}

\end{document}